%% file: text.tex
\documentclass[prl,twocolumn,preprintnumbers,footinbib,tightenlines,superscriptaddress]{revtex4}

\pdfoutput=1
\usepackage[english]{babel}
\makeatletter\AtBeginDocument{\let\@elt\relax}\makeatother 
\usepackage{amsmath,amssymb,amsfonts, bm,bbm,slashed, subdepth}
\usepackage{graphicx}
\usepackage[sort&compress]{natbib}
\usepackage{xcolor}
\usepackage{hyperref}
\usepackage{cleveref}
\usepackage{enumerate}
\usepackage{epsfig, subfigure}
\usepackage{setspace}
\usepackage{booktabs, tabularx}
\usepackage{units}
\usepackage{placeins}
\usepackage{multirow}
\usepackage{mathtools}

\newcommand{\z}{\ell}
\newcommand{\omegapl}{\omega_\text{pl}}
\newcommand{\BB}{\text{eq}}
\newcommand{\ini}{\text{ini}}
\newcommand{\EQ}{\text{dec}}
\newcommand{\RI}{\text{rei}}
\newcommand{\intf}{{\cal P}}
\newcommand{\losc}{\z_\text{osc}}

\begin{document}

\preprint{DESY 20-097}
\preprint{CERN-TH-2020-082}

\title{Potential of radio telescopes as high-frequency gravitational wave detectors}


\author{Valerie Domcke}
\affiliation{Deutsches Elektronen-Synchrotron DESY, Notkestrasse 85,
22607 Hamburg, Germany}
\affiliation{Theoretical Physics Department, CERN,
1 Esplanade des Particules, CH-1211 Geneva 23, Switzerland}
\affiliation{Institute of Physics, Laboratory for Particle Physics and Cosmology (LPPC),
\'Ecole Polytechnique F\'ed\'erale de Lausanne (EPFL), CH-1015 Lausanne, Switzerland}

\author{Camilo Garcia-Cely}
\affiliation{Deutsches Elektronen-Synchrotron DESY, Notkestrasse 85,
22607 Hamburg, Germany}

\begin{abstract}
In the presence of magnetic fields, gravitational waves are converted into photons and vice versa.  
We demonstrate that this conversion leads to a distortion of the cosmic microwave background (CMB), which can serve as a detector for MHz to GHz gravitational wave sources active before reionization. 
The measurements of the radio telescope EDGES can be cast as a bound on the gravitational wave amplitude, $h_c < 10^{-21} (10^{-12})$ at 78 MHz, for the strongest (weakest) cosmic magnetic fields allowed by current astrophysical and cosmological constraints. Similarly, the results of ARCADE 2 imply $h_c < 10^{-24} (10^{-14})$ at $3 - 30$~GHz. For the strongest magnetic fields, these constraints exceed current laboratory constraints by about seven orders of magnitude.
Future advances in 21cm astronomy may conceivably push these bounds below the sensitivity of cosmological constraints on the total energy density of gravitational waves. 
\end{abstract}

\maketitle
Gravitational waves (GWs) produced in the early Universe~\cite{Maggiore:2018sht,Caprini:2018mtu} can traverse cosmic distances without experiencing any interactions, making them a unique probe of very high energy physics.
Since the comoving Hubble horizon grows with time, GWs produced at energies around the scale of grand unification 
have frequencies in the MHz and GHz regime today, far beyond the reach of the laser interferometers LIGO, VIRGO or KAGRA. See~\cite{REECE1984341,Cruise:2006zt,Akutsu:2008qv,Ito:2020wxi,Cruise_2012,Ejlli:2019bqj} for some existing laboratory bounds at these frequencies. 

Here we focus on searching for high-frequency GWs exploiting the (inverse) Gertsenshtein effect~\cite{Gertsenshtein,Boccaletti1970}, which describes the conversion of GWs into photons in the presence of a magnetic field (see e.g.~\cite{DeLogi:1977qe,Raffelt:1987im,Macedo:1984di,Fargion:1995mm, Dolgov:2012be,Cruise_2012,Ejlli:2019bqj,Ejlli:2020fpt}).
As an immediate consequence of general relativity and classical electromagnetism, this is a purely SM process.  Involving gravity, the conversion probability is extremely small which may, however, be compensated by considering a `detector' of cosmological size. 
In fact, magnetic fields with cosmological correlation lengths might well permeate our Universe with certain astrophysical observations strongly suggesting a lower limit of order $\unit[10^{-16}]{G}$~\cite{Neronov:1900zz,Tavecchio:2010mk,Takahashi:2013lba}, and the CMB setting an upper bound in the pG-nG range~\cite{Jedamzik:2018itu,Ade:2015cva,Pshirkov:2015tua}. See~\cite{Durrer:2013pga} for a comprehensive review.  

The pioneering study~\cite{PhysRevLett.74.634} proposed the inverse Gertsenshtein effect in cosmic magnetic fields to search for GWs but neglected the plasma mass of photons, as pointed out in Ref.~\cite{Cillis:1996qy}. The idea was revisited in Ref.~\cite{Pshirkov:2009sf} suggesting an observable effect, however as noted in \cite{Dolgov:2012be} decoherence effects were not correctly accounted for. 
More recently~\cite{Fujita:2020rdx} studied the production of GWs from CMB photons. In this letter, we focus on CMB distortions arising from the Gertsenshtein effect during the dark ages, i.e.\ the period between recombination and reionization. Due to the small fraction of free electrons in this period, the effective plasma mass of the photons is suppressed, increasing the conversion probability between GWs and photons. Taking into account inhomogeneities in the thermal plasma and in the cosmic magnetic fields, we demonstrate that existing measurements of the Rayleigh-Jeans tail of the CMB spectrum, performed e.g. by ARCADE 2~\cite{Fixsen_2011} and by EDGES~\cite{Bowman:2018yin}, can be translated into constraints on GWs in the MHz-GHz regime. These are competitive with, or even exceed, current laboratory constraints, depending on the assumptions on the cosmic magnetic fields.

\begin{figure}[t]
\includegraphics[width =0.69\columnwidth]{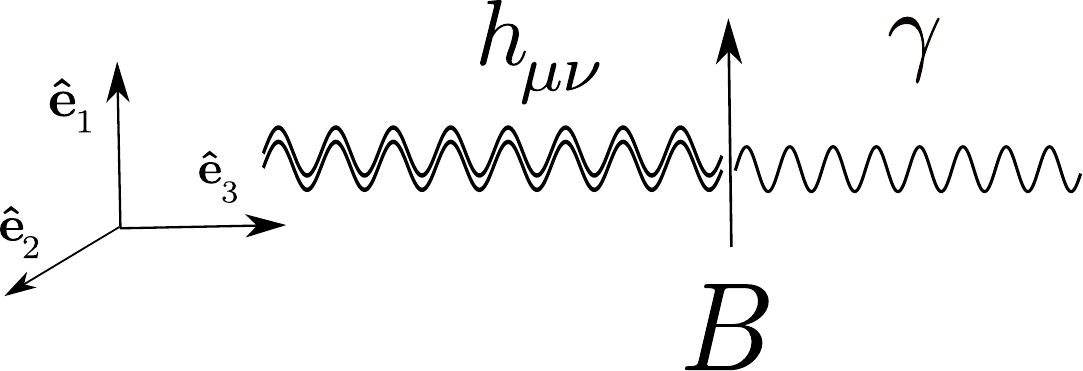}
\caption{
The Gertsenshtein effect.
 }
\label{fig:sketch}
\end{figure}

\section{The Gertsenshtein effect}

Calculating the conversion rate for this oscillation process requires solving Maxwell's equations for the vector potential, $A^\mu$, describing the electromagnetic radiation, together with the linearized Einstein's equations for the metric $g_{\mu\nu} =\eta_{\mu\nu}+h_{\mu\nu}$, in which $h_{\mu\nu}$ describes the GWs. In this work we will adopt $\eta_{\mu\nu} = \text{diag}(+---)$ and work with natural Heaviside-Lorentz units ($\hbar=c=1$), except in this section where we keep fundamental constants explicitly   to emphasize that the Gertsenshtein effect is a classical phenomenon.

Let us ignore the Universe expansion first and consider a GW propagating in the $\mathbf{\hat{e}_3}$ direction inside a fixed box of size $\Delta \z$ that contains a uniform  transverse  magnetic field, $B$, and a non-negligible  uniform density of free electrons, $n_e$. Without loss of generality, we assume that the magnetic field points in the $\mathbf{\hat{e}_1}$ direction. See Fig.~\ref{fig:sketch}.  In this coordinate system we introduce $h_\times = h^{12} =h^{21}$ and $A_\times = A^1$ as well as $h_+ = -h^{22} = h^{11}$ and $A_+ =-A^2$. This is because the aforementioned equations can be elegantly cast as~\cite{Dolgov:2012be,Raffelt:1987im}~\footnote{The first relation in Eqs.~\eqref{eq:waveeq} differs from that of~\cite{Fujita:2020rdx} in the negative sign.}  
\begin{align}
\left(\Box+\omega^2_\text{pl}/c^2 \right) A_\lambda =- B \partial_\z h_\lambda\,,&&
\Box h_\lambda = \kappa^2 B \partial_\z A_\lambda\,,
\label{eq:waveeq}
\end{align}
where $\lambda \in \{+,\times\}$,  $\z$  is the third component,   $\Box =\partial_t^2/c^2-\partial_\z^2$, $\kappa =\left(16 \pi G\right)^{1/2}/c^2$. 
We include the plasma frequency $\omega_\text{pl}= \sqrt{e^2n_e/m_e}$, which acts as an effective mass term and gives electromagnetic  waves of frequency $\omega$ a refractive index $\mu = \sqrt{1-\omegapl^2/\omega^2}$ when $B\to 0$.  
Eq.~\eqref{eq:waveeq} also applies for arbitrary uniform fields with $B$ interpreted as the corresponding transverse component.
See the supplementary material for more details.
Assuming a plane wave traveling in the positive direction with $\omega \geq \omegapl$, 
the exact solution  of Eqs.~\eqref{eq:waveeq} (see also~\cite{Ejlli:2020fpt}) can be written as 
\begin{align}
\hspace{-8pt}
\psi(t,\z)  \equiv
\begin{pmatrix}
\sqrt{\mu}\,\, A_\lambda \\
\frac{1}{\kappa}\, h_\lambda
\end{pmatrix}
=
e^{-i\omega t}
e^{i K \z}
\psi(0,0)\,,\,
\label{eq:sol}
\end{align}
with $K$ being the Hermitian matrix
\begin{align}
K = 
\begin{pmatrix}
 \frac{ \mu}{c}\sqrt{\omega^2+\left(\frac{\kappa B}{1+\mu}\right)^2} \,  &  -i\frac{\sqrt\mu\,\kappa B }{1+\mu}\\
i \frac{\sqrt\mu\, \kappa B }{1+\mu} &  \frac{1}{c}\sqrt{\omega^2+\left(\frac{\kappa B}{1+\mu}\right)^2} 
\end{pmatrix}
 \,.
\label{eq:U}
\end{align}

It is convenient to introduce $\psi$ because its magnitude, $|\psi(t,\z)|^2$ , is 
conserved. 
This 
 easily follows from the unitarity of the matrix ${\cal U(\z)}=e^{i K \z} $. 
In particular, $\psi(0,0)=(0,h_{\lambda,0}/\kappa)$ for a pure GW state entering the box, and consequently $ \psi(t,\Delta \z) = e^{-i\omega t}\left({\cal U}_{12} (\Delta\z)  ,{\cal U}_{22} (\Delta\z)\right) h_{\lambda,0}/\kappa$ after leaving the box. Since $|{\cal U}_{12} (\Delta\z)|^2 + |{\cal U}_{22} (\Delta\z)|^2 =1$,  the quantity $P(\Delta \z)\equiv |{\cal U}_{12} (\Delta \z)  |^2$ can be interpreted  as the probability of GW conversion after traversing a distance $\Delta \z$. Simple algebra shows  
\begin{equation}
 P (\Delta \z) =|K_{12}|^2 \,\losc^2\, \sin^2(\Delta\z/\losc)\,,
\label{eq:P0}
\end{equation}
with $\losc^{-1}  = 
\sqrt{\omega^2(1-\mu)^2/c^2+ \kappa^2 B^2}/2$. These expressions reduce to the approximated formulae previously found
(see e.g.~\cite{Raffelt:1987im,Ejlli:2018hke}).

Although cosmic magnetic fields are not expected to be perfectly homogeneous, coherent oscillations take place in highly homogeneous patches, for which $\losc \ll \Delta \z$ and therefore $P(\Delta \z)  = |K_{12}|^2 \losc^2/2$ on average.  Taking into account inhomogeneities in $n_e$~\footnote{See the discussion in the Supplemental Material which includes Refs.~\cite{Venhlovska:2008uc,Dvorkin:2013cga,2012PhRvD..85d3522F, Carlson:1994yqa} } and $B$, the coherence of the $g \leftrightarrow \gamma$ oscillations is lost on distances larger than $\Delta \z$, that is, the smallest distance  on which $B$  and $n_e$ are uniform. Denoting the total distance traveled by the GW as $D$, this corresponds to traversing $N = D/\Delta \z$ independent regions with a conversion probability $P(\Delta \z)$ each. As long as $N \cdot P(\Delta \z) \ll 1$, 
this gives a total conversion probability of  $P(D) = D |K_{12}|^2 \losc^2/(2\Delta \z)$
~\cite{Cillis:1996qy,Pshirkov:2009sf},  corresponding to an average conversion rate (i.e. probability per time)~\footnote{See the discussion in the Supplemental Material showing that the effect of the magnetic field on the wave velocity is negligible, which includes a reference to~\cite{Monitor:2017mdv} } given by
\begin{equation}
\langle\Gamma_{g\leftrightarrow\gamma}\rangle =\frac{c\,|K_{12}|^2 \losc^2}{2   \Delta \z}\,.
\label{eq:rate}
\end{equation}
In the supplementary material we demonstrate that this simple estimate correctly captures the essential features of a more involved computation based on the expected power spectrum of the magnetic field. Note that any additional inhomogeneities would further enhance the conversion rate by limiting the coherence of the $g \leftrightarrow \gamma$ oscillations.

\begin{figure*}[t]
\includegraphics[height =0.365\textheight]{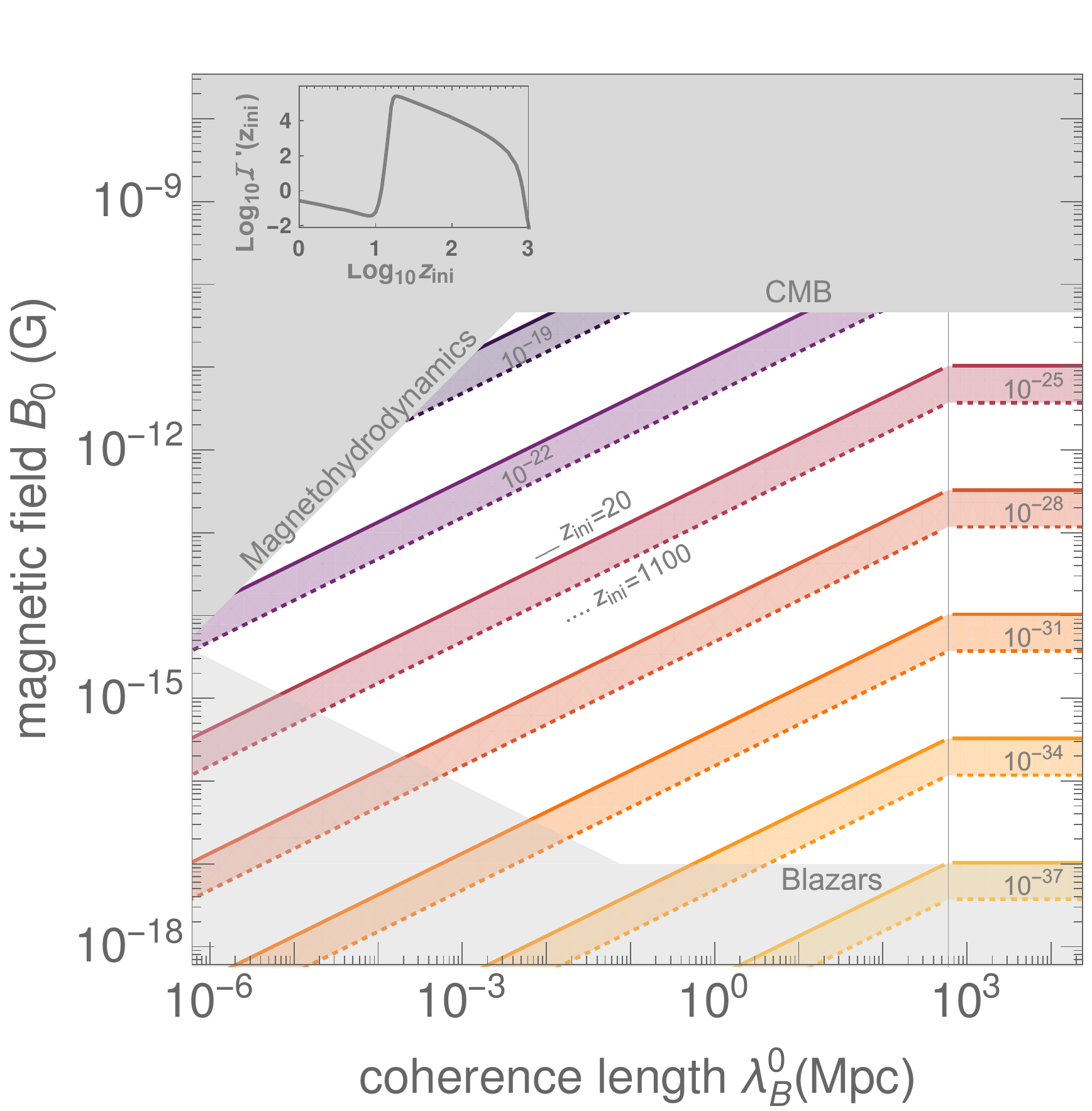}\hspace{10pt}
\includegraphics[height =0.365\textheight]{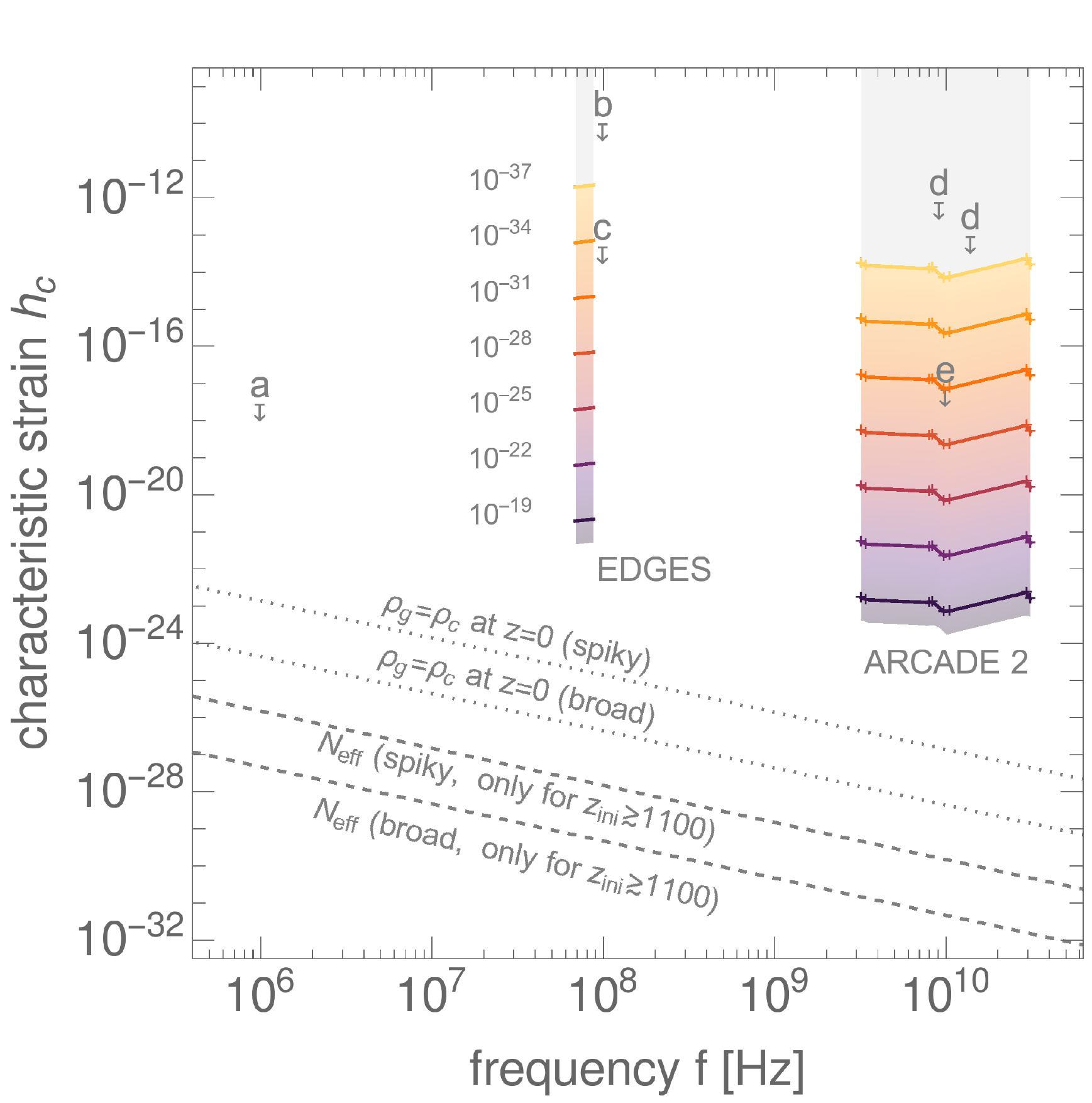}
\caption{
\emph{Left:} 
Parameter space for cosmic magnetic fields today. Gray shaded areas show the exclusion discussed in the text. 
The solid (dashed) colored curves indicate contour lines for the re-scaled conversion probability, $(T_0/\omega_0)^2\intf$. See Eqs.~\eqref{eq:integratedP} and \eqref{eq:intf}. 
\emph{Right:} 
Upper bounds on the stochastic GW background derived from ARCADE2 and EDGES (this work), compared to existing laboratory bounds from a) superconducting parametric converter~\cite{REECE1984341}, b) waveguide~\cite{Cruise:2006zt}, c) 0.75 m interferometer~\cite{Akutsu:2008qv}, d) magnon detector~\cite{Ito:2020wxi} and e) magnetic conversion detector~\cite{Cruise_2012}. The solid lines indicate the allowed parameter space for cosmic magnetic fields, as given in the left panel. The dashed lines mark the $N_\text{eff}$ constraint for broad GW spectra and for a peaked spectrum with $\Delta \omega/\omega = 10^{-3}$. For reference, the dotted lines indicate $\rho_{g} = \rho_c$.
}
\label{fig:hc}
\end{figure*}

We now include the effect of the Universe expansion during the dark ages.  This is the period between photon decoupling and reionization, 
$z_\EQ  \simeq 1100 \gtrsim z \gtrsim  z_\RI \simeq 10$, beginning with the formation of the CMB and ending when the first stars were formed. During this time,  
the refractive index of MHz-GHz CMB photons is determined by the tiny electron density, with the contributions of neutral hydrogen, helium and birefringence being subdominant~\cite{Chen:2013gva,Kunze:2015noa,Mirizzi:2009iz}.  This allows us to adopt Eq~\eqref{eq:rate}, after a few modifications. The conversion probability in an adiabatic expanding Universe
is simply the line-of-sight integral of the rate
\begin{equation}
\intf \equiv \int_{l.o.s.} \langle \Gamma_{g\leftrightarrow \gamma}  \rangle dt = \int^{z_\ini}_{0} \frac{\langle\Gamma_{g\leftrightarrow \gamma}\rangle}{ (1+z)\,H} dz \,,
\label{eq:integratedP}
\end{equation} 
where we use null-geodesics $Hdt= dT/T =dz/(1+z) $. Also,  $z_\ini\leq z_\EQ $ is an initial condition to be specified below and $H =  H_\EQ \left(T/T_\EQ\right)^{3/2}$ is the Hubble parameter during the dark ages, which are matter dominated.  
Furthermore, the average magnetic energy density of the Universe $\rho_B= B^2/2$ redshifts as $\rho_B=\rho_{B0}\left(1+z\right)^4$~\footnote{In Eq.~\eqref{eq:rate}, $B^2 \to 2 B^2/3$ extracts the transverse component.}.
Additionally, such a field is associated with a coherence length, $\lambda_B=\lambda^0_B/(1+z)$, because it is not expected to be homogeneous everywhere.  
Concerning these two quantities we emphasize three important facts here and refer the reader to \cite{Durrer:2013pga} for a more comprehensive discussion: i) a recent CMB analysis gives  $B_0\lesssim \unit[47]{pG}$~\cite{Jedamzik:2018itu}  ii) Blazar observations strongly suggest a lower limit on $B_0$~\cite{Caprini:2015gga}  because otherwise their gamma-ray spectra can not be explained under standard cosmological assumptions~\cite{Neronov:1900zz,Tavecchio:2010mk,Takahashi:2013lba,Chen:2014rsa}, and iii)  magnetohydrodynamic turbulence damps out  large magnetic fields at small distances, imposing an additional (theoretical) upper limit~\cite{Durrer:2013pga}.   Fig.~\ref{fig:hc} show these constraints. 

In addition, the electron number density during this epoch is $n_e(z) = n_{b0} (1+z)^3 X_e(z)
$, where $n_{b0}=\unit[0.251]{m^{-3}}$  is  the baryon number density today~\cite{Aghanim:2018eyx} and $X_e(z)$ is the ionization fraction, taking values $1,\, 0.68,\,0.0002$ and $0.15$  at $z=0,\,10,\,20$ and $1100$, respectively~\footnote{For numerical computations, we use the values reported in Ref.~\cite{Kunze:2015noa}}. This gives  plasma frequencies today, $\omega_{\text{pl},0}$, lying in the Hz range, which allows us to take $1-\mu(z) =(1+z)X_e(z)\omega_{\text{pl},0}^2/(2\omega_0^2)\ll 1$, for waves of frequency $\omega= \omega_0 (1+z)$ with $\omega_0 \sim$~GHz.
 Moreover, $B_0\lesssim \unit[47]{pG}$ results in the oscillation length being numerically dominated by the plasma frequency so that 
$
\losc^{-1} = (1+z)^2 X_e(z)\omega_{\text{pl},0}^2/(4\omega_0c)
$.
This gives $\losc\ll\unit[1]{pc}\ll \Delta \z$, as anticipated above. 
Here, in order to account for electron inhomogeneities we conservatively take $\Delta \z = \Delta \z_0/(1 + z)$ to be given by $\Delta \z_0 = \text{min}[\lambda_\text{EQ}, \lambda_B^0]$ where $\lambda_{EQ}/(2\pi) = \unit[95]{Mpc}$ is the characteristic comoving scale for the onset of structure formation (corresponding to the perturbation mode entering the horizon at matter-radiation equality).  
Putting all this together, we obtain 
\begin{eqnarray}
\intf  \simeq \,6.3\times 10^{-19} \left( \frac{B_0}{\text{nG}} \right)^2
 \left( \frac{\omega_0}{T_0} \right)^2
\hspace{-3pt}
 \left( \frac{\text{Mpc}}{\Delta \z_0}\right)  
\left( \frac{{\cal I}(z_\ini)}{ 10^6}\right),
\label{eq:intf}
\end{eqnarray} 
with $T_0/(2 \pi) = 2.725~\text{K}/(2 \pi) = 56.78$~GHz and  ${\cal I}(z_\ini)=\int_0^{z_\ini}dz\,(1+z)^{-3/2}X_e^{-2}(z)$.
The left panel of Fig.~\ref{fig:hc} displays contours of $(T_0/\omega_0)^2\intf$ in the parameter space of cosmic magnetic fields. The inset shows ${\cal I}'(z_\ini)$, explaining the weak redshift-dependence of ${\cal I}(z_\ini)$, with the largest contribution arising from $z\sim 10$.

\section{CMB Distortions}

The CMB photon distribution, $f_\gamma (\omega,T)$,  retains its equilibrium form 
during the  dark ages, i.e. is given by a black-body spectrum, $f_\BB = 1/(e^{\omega/T}- 1)$ with $\omega/T=\omega_0/T_0$. Our aim here is to calculate deviations  from such a spectrum, $\delta f_\gamma = f_{\gamma} -f_\BB$.

The spectrum of GWs is  commonly characterized by $\Omega_\text{GW}$, which parametrizes the corresponding energy density per logarithmic frequency bin. This quantity can be used to introduce -- in an analogous manner to $f_\gamma$-- the distribution function for GWs, $f_g$. More precisely, in terms of it, the energy density is given by 
\begin{eqnarray}
 \rho_{g}(T) \hspace{-2pt} =\hspace{-2pt} \int \frac{d\ln\omega}{ \pi^2} \, \omega^4 f_g  \equiv \rho_c(T) \hspace{-2pt}\int d \ln\omega \,\, \Omega_\text{GW}\left(\frac{\omega}{2 \pi},T\right) \label{eq:rhog2},
\end{eqnarray}
with $\rho_c(T)$ denoting the Universe total energy density. 

Both distributions satisfy the Boltzmann equation $ \hat{L} f_{\gamma/g} =\pm \langle\Gamma_{g\leftrightarrow\gamma}\rangle (f_g -f_\gamma)$, where  $\hat{L} \equiv\partial_t - H \omega \partial_\omega = - H \left( T \partial_T + \omega \partial_\omega \right)$ is the corresponding Lioville operator. Its solution leads to 
\begin{eqnarray}
\delta f_\gamma (\omega_0,T_0)
=\left( f_g (\omega_\text{ini},T_\text{ini})- f_\BB 
\right)
\intf+ {\cal O}(\intf^2)\,,
\label{eq:master}
\end{eqnarray}
with $\intf$ defined as in Eq.~\eqref{eq:integratedP}. 
We solve the Boltzmann equations from an initial temperature $T=T_\ini$ -- when the photon distribution is a black-body spectrum, i.e $f_\gamma(\omega,T_\ini)= f_\BB (\omega/T_\ini)$ --  until today. 
If decoupling is prior to  the GW emission, the latter fixes $T_\ini$. Otherwise, we set $T_\text{ini} =T_\text{dec}$ because the ionization fraction sharply drops after $z \sim z_\EQ$ rendering any prior contribution negligible. This is illustrated in the inset of Fig.~\ref{fig:hc} (left panel), which also shows that the conversion rate is anyways largely insensitive to the  precise value of $T_\ini$.

Eq.~\eqref{eq:master} can alternatively be derived by  considering the density-matrix formalism. In that case case, $f_\gamma$ and $f_g$ are proportional to the diagonal entries of such a matrix, which evolves by means of the Hamiltonian associated with Eq.~\eqref{eq:U}. See the supplementary material for details. The fact that  using both methods we obtain the same result --i.e. Eq.~\eqref{eq:master}-- is reassuring and indicates that  decoherence effects are properly taken into account~\cite{Dolgov:2012be}.  Due to this as well as the way we treat inhomogeneities, our results differ from those of~\cite{Pshirkov:2009sf}.

\section{Constraints on the stochastic GW background}

In this letter we focus on the Rayleigh-Jeans part of the CMB, i.e. $\omega \ll T$ implying $f_\text{eq} \simeq T/\omega$. In this regime, a subdominant GW contribution to the total radiation energy density is compatible with $f_g \gg f_\gamma~$\footnote{See Ref.~\cite{Pospelov:2018kdh} for a related argument for decaying dark matter.}, and can thus produce an enhancement of the low-frequency CMB tail through the first term of Eq.~\eqref{eq:master}. More precisely, the assumption $f_g > f_\gamma$ translates to $\Omega_\text{GW}/\Omega_\gamma > 15/\pi^4 (\omega/T)^3$ as can be seen by rewriting Eq.~\eqref{eq:rhog2} as
$
 \Omega_\text{GW} =\omega^4  f_g\left(\omega,T\right)/(\pi^2 \rho_c) = (15/\pi^4) \, \left(\omega/T\right)^4 f_g(\omega,T) \,  \Omega_\gamma  \,
$
with  $\Omega_\gamma = \rho_\gamma/\rho_c =   \pi^2 T^4/(15\,\rho_c)$. Even a scale-invariant GW spectrum as small as $\Omega_\text{GW} \simeq 10^{-15}$ implies $f_g > f_\gamma$ at e.g.\ $\omega/T \simeq 10^{-3}$.

With $\omega\ll T$ and $f_g\gg f_\gamma$, Eq.~\eqref{eq:master} reads

\begin{equation}
  \frac{\delta f_\gamma }{f_\gamma} \left(\omega_0,T_0\right)= \frac{\pi^4}{15}\left(\frac{T}{\omega}\right)^3\intf \, \frac{\Omega_\text{GW}}{\Omega_\gamma}\,. 
 \label{eq:pred}
\end{equation}

For a given detector sensitivity $\delta f_\gamma/f_\gamma$ and a given value of the conversion probability $\intf$, relation~\eqref{eq:pred} sets stringent bounds on the GW spectrum, which can be expressed in terms of the characteristic strain by means of~\cite{Maggiore:1900zz}
\begin{equation}
 h_c = \left(\frac{3 H_0^2}{4 \pi^2} \Omega_\text{GW} f^{-2} \right)^{1/2}\,.
\end{equation}
This is related to the one-sided power spectral density $S_h$ as $h_c = \sqrt{f S_h(f)}$. 
Fig.~\ref{fig:hc} contrasts the resulting constraints with existing bounds in the literature.

\paragraph{$N_\text{eff}$ bound.} GWs contribute to the energy budget of the Universe in the form of radiation and are as such constrained by the BBN and CMB bounds on the effective number of massless degrees of freedom $N_\text{eff}$~\cite{Pagano:2015hma},
\begin{align}
 \rho_{g}(T)   \leq \frac{7}{8} \left( \frac{4}{11} \right)^{4/3} \Delta N_\text{eff} \, \rho_\gamma(T) \,,
 \label{eq:Neffbound}
\end{align}
with $\Delta N_\text{eff} \lesssim 0.1$~\cite{Cyburt:2015mya,Aghanim:2018eyx}.  For a spectrum $\Omega_\text{GW}$ which is approximately scale invariant between $f_\text{min}$ and $f_\text{max}$ with $\ln(f_\text{max}/f_\text{min}) \sim {\cal O}(1)$,
this implies
\begin{equation}
 \frac{\Omega_\text{GW}}{\Omega_\gamma} \lesssim \frac{7}{8} \left( \frac{4}{11} \right)^{4/3} \Delta N_\text{eff}\,,
 \label{eq:Neff}
\end{equation}
whereas for a narrow spectrum peaked at $\tilde \omega$ with width $\Delta \tilde \omega \lesssim \tilde \omega$ this bound is relaxed by a factor $(\tilde \omega/\Delta \tilde \omega)$. Note that this bound applies only to GWs present already at  
CMB decoupling.

\paragraph{Probing the Rayleigh-Jeans tail of the CMB spectrum.} Below $\omega_0/T_0 \sim 10^{-2}$, galactic foregrounds dominate the radio sky. Here we focus on the results reported by ARCADE2~\cite{Fixsen_2011} which covers the sweet spot of the low-frequency Rayleigh-Jeans spectrum before galactic foregrounds become important ($f = \omega_0/(2 \pi) =  3$, 8, 10, 30 and 90 GHz) and by EDGES~\cite{Bowman:2018yin}, which is a recent measurement of the global 21cm absorption signal at 78~MHz. 

ARCADE2 was a balloon experiment equipped with a radio receiver measuring the black body temperature of sky~\cite{Fixsen_2011}. The cleanest frequency band is around 10~GHz enabling a mK resolution, $\delta f_\gamma/f_\gamma = \delta T/T_\text{CMB} \lesssim 4 \times 10^{-4}$ at $\omega_0/T_0  \simeq 0.18$. At smaller frequencies, ARCADE2 observed a significant radio excess beyond the expected galactic foreground whose origin remains an open question (see e.g.~\cite{Seiffert_2011,Feng:2018rje}). Assuming that this excess is entirely astrophysical, we can impose an upper bound on an additional contribution from a stochastic GW background using the 3, 8, 10 and 30~GHz frequency bands. In Fig~\ref{fig:hc}, these frequencies are marked by crosses, the solid lines connecting them serve only to guide the eye.

Recently, the first observation of the global (i.e.\ sky-averaged) 21cm absorption signal was reported by the EDGES collaboration~\cite{Bowman:2018yin}. The absorption feature was found to be roughly twice as strong as previously expected, which if true, would indicate that either the primordial gas was significantly colder or the radiation background was significantly hotter than expected. Conservatively, we may assume that the deviation from the expected value is due to foreground contamination, and place a bound on any stochastic GW background by using $\delta f_\gamma / f_\gamma \lesssim 1$ at $\omega/T = 1.4 \times 10^{-3}$ (78 MHz). The width of the observed absorption feature (19~MHz) determines the width of the frequency coverage.

\section{Discussion}

Cosmological sources of GWs typically produce stochastic GW backgrounds with a frequency roughly related to the comoving Hubble horizon at the time of production. Processes in the very early Universe at energy scales far beyond the reach of colliders thus generically produce GWs in the MHz and GHz regime. Despite the large amount of redshift, these violent processes can produce sizeable GW signals, saturating the $N_\text{eff}$ bound~\eqref{eq:Neffbound}. Some examples are axion inflation~\cite{Barnaby:2010vf}\footnote{The dominant GW signal of axion inflation may also be related to its subsequent preheating phase~\cite{Adshead:2019lbr}, also leading to MHz to GHz signal which can easily saturate the $N_\text{eff}$ bound.}, metastable cosmic strings~\cite{Buchmuller:2019gfy} and evaporating light primordial black holes~\cite{Anantua:2008am,Dong:2015yjs}.  Further significant contributions may be expected from preheating~\cite{Dolgov:1989us,Traschen:1990sw,Kofman:1994rk,Kofman:1997yn,Figueroa:2017vfa} and first order phase transitions occurring above $10^7$~GeV~\cite{Witten:1984rs,Hogan:1986qda,Punturo:2010zz,Hild:2010id,Evans:2016mbw}. The sensitivity of radio telescopes can however not yet compete with the cosmological $N_\text{eff}$ bound, unless one considers essentially monochromatic signals (which may arise e.g. from large monochromatic scalar perturbations~\cite{Bugaev_2010, Saito_2010}).

Since the dominant contribution to $\intf$ arises around reionization, a particularly interesting target are GW sources active around $10 \lesssim z \lesssim  10^3$, which would not be constrained by the $N_\text{eff}$ bound. During the dark ages, there is no generic reason to expect GW production in the GHz regime but there are models which predict such a signal for suitable parameter choices. For example, mergers of light primordial black holes in this epoch (with masses of about $10^{-9..-7}~M_\odot$) would result in GHz GW signals today~\cite{Maggiore:1900zz}, see~\cite{Raidal_2019,Gow_2020} for a discussion of possible rates. Superradiant axion clouds around spinning black holes yield an essentially monochromatic GW signal with $f \lesssim$~MHz~\cite{Arvanitaki:2009fg,Arvanitaki:2010sy,Arvanitaki:2012cn}, with higher frequency possible when considering primordial black holes with masses below the Chandrasekhar limit. 

We emphasize that the use of radio telescopes allows to search for GWs in a wide frequency regime. While the absence of any excess radiation can already constrain some models under the assumption of strong cosmic magnetic fields, the potential of this method will truly unfold with further improvements in the sensitivity of radio telescopes --driven in particular by the advances in 21cm cosmology-- or in the case of a positive detection of excess radiation. 

An example of future advances in radio astronomy is the case of the Square Kilometer Array (SKA). Assuming an effective area per antenna temperature of at least $\unit[10^2]{m^2/K}$~\cite{Braun:2019gdo}\footnote{See also \url{https://www.skatelescope.org/wp-content/uploads/2014/03/SKA-TEL_SCI-SKO-SRQ-001-1_Level_0_Requirements-1.pdf}} in the $0.1-10$~GHz range, a few hours of observation will lead to sensitivities in the ballpark of $\unit[]{\mu Jy}$, which must be compared against CMB fluxes of at least $\unit[10^3]{Jy}$. SKA measurements are thus very promising although sufficient foreground subtraction will be extremely challenging.

\paragraph{Acknowledgments.} It is a pleasure to thank Nancy Aggarwal, Sebastien Clesse, Mike Cruise, Hartmut Grote and Francesco Muia for insightful discussions on the Gertsenshtein effect and high-frequency GW sources at the ICTP workshop ``Challenges and opportunities of high-frequency gravitational wave detection''.
Likewise, we would also like to thank Torsten Bringmann,  Damian Ejlli, Kohei Kamada and Kai Schmidt-Hoberg. 
This work was partially funded by the Deutsche Forschungsgemeinschaft under Germany's Excellence Strategy - EXC 2121 ``Quantum Universe'' - 390833306.  C.G.C. is supported by the
Alexander von Humboldt Foundation.

\input{Supplementary.tex}

\bibliographystyle{JHEP}
\bibliography{ref}

\end{document}

%% file: Supplementary.tex
\begin{widetext}
\section{Supplementary material}

\subsection{ The wave equation}
\paragraph{Gravitational radiation sourcing electromagnetic waves. } Maxwell's equations in curved space-time can be written as 
\begin{equation}
0=\nabla_{\nu} F^{\mu \nu }+\frac{1}{c}j^\mu=  \frac{1}{\sqrt{-g}} \partial_\nu \left( \sqrt{-g}F^{\mu\nu}\right)+ \frac{1}{c}j^\mu\,,
\label{eq:Maxwell0}
\end{equation}
where $j^\mu$ and $F^{\mu\nu}$ are respectively the electromagnetic current and strength-field tensor in  Lorentz-Heaviside units. In this work $\eta_{\mu\nu} = \text{diag}(+ - - -)$. 
For metric fluctuations describing a gravitational wave, $g_{\mu\nu} =\eta_{\mu\nu}+h_{\mu\nu}$, we have $\partial_\nu  \sqrt{-g}/\sqrt{-g} = g^{\rho\sigma} \partial_\nu g_{\rho\sigma}/2 =  \partial_\nu \left( \eta^{\rho\sigma}h_{\rho\sigma}  \right)/2 + {\cal O}(h^2)$. 
Moreover, for an electromagnetic wave determined by the vector potential, $A^{\mu}$, and propagating in the presence of a uniform and constant magnetic field, $F^\text{ext}_{\mu\nu}$, the part of the electromagnetic tensor linear in only $A$ or $h$ is
\begin{equation}
 F^{\mu\nu}\supseteq \left(\eta^{\mu\lambda} \partial_{\lambda} A^{\nu}-\eta^{\nu\lambda} \partial_{\lambda} A^{\mu}\right)+ \left(\eta^{\alpha\mu}\eta^{\beta\nu}-h^{\alpha\mu}\eta^{\beta\nu}-\eta^{\alpha\mu}h^{\beta\nu} \right) F^{\text{ext}}_{\alpha\beta}\,.
\label{eq:F}
\end{equation}
Notice that $F^{\text{ext}\,{\alpha\beta}}$ varies due to the fluctuating metric in contrast to $F^{\text{ext}}_{\alpha\beta}$, which does not change. Then, Eq.~\eqref{eq:Maxwell0} to leading order is
\begin{equation}
 \eta^{\mu\lambda} \partial_{\lambda} \partial_\nu A^{\nu}-\eta^{\nu\lambda} \partial_\nu\partial_{\lambda} A^{\mu}-\partial_\nu h^{\alpha\mu}\eta^{\beta\nu}F^{\text{ext}}_{\alpha\beta}-\eta^{\alpha\mu}\partial_\nu h^{\beta\nu}  F^{\text{ext}}_{\alpha\beta}  + \frac{1}{2}  \partial_\nu  \left(\eta^{\rho\sigma} h_{\rho\sigma}  \right)F^{\text{ext}}_{\alpha\beta} \eta^{\mu\alpha} \eta^{\nu\beta}+\frac{1}{c}j^\mu=0\,.
\label{eq:Maxwell1}
\end{equation}
In this work we will adopt the harmonic-Lorentz gauge:
\begin{align}
\partial_\nu h^{\beta\nu}  = \frac{1}{2} \eta^{\beta\nu} \partial_\nu \left( \eta^{\rho\sigma} h_{\rho\sigma} \right)&&\text{and}&& \partial_\nu A^\nu =0 \,,
\end{align}
which simplifies the previous equation to 
\begin{equation}
 -\eta^{\nu\lambda} \partial_\nu\partial_{\lambda} A^{\mu}-\partial_\nu h^{\alpha\mu}\eta^{\beta\nu}F^{\text{ext}}_{\alpha\beta} +\frac{1}{c}j^\mu =0\,.
\label{eq:Maxwell2}
\end{equation}

During the dark ages, a small fraction of electrons are free and can gain momentum from the electromagnetic wave creating a non-vanishing current $j_\mu$. More precisely, such momentum is given by the Lorentz force as $p^ i =  (e/c)\int  F^i_{\,\,\,\nu}dx^\nu$, which is approximately $ p^i\simeq -e A^i/c $ up to relativistic corrections.
The corresponding velocity induces a current $j^i =e n_e p^i/m_e$. Therefore
\begin{align}
j^\mu = - \frac{1}{c} \omega_\text{pl}^2 A^\mu  \,, &&\text{where}&& 
\omega_\text{pl}\equiv \sqrt{\frac{e^2n_e}{m_e}}\,.
\end{align}
For waves traveling in the $3$-direction, $\partial_i = (0,0,\partial_{\z})$. Hence, Eq.~\eqref{eq:Maxwell2} for the transverse components can be cast as

\begin{equation}
\left(\Box+\omega^2_\text{pl}/c^2 \right)\begin{pmatrix}A^1 \\ A^2\end{pmatrix} = 
\begin{pmatrix}
\partial_{\z} h^{11} & \partial_{\z} h^{12}\\
\partial_{\z} h^{12} & \partial_{\z} h^{22}
\end{pmatrix}
\begin{pmatrix}
B^2\\
-B^1\\
\end{pmatrix}\,,
\label{eq:Aeq0}
\end{equation}
where, as usual, $F^\text{ext}_{23} =-B^1$, $F^\text{ext}_{13} =B^2$ and  $\Box =\partial_0^2/c^2-\partial_{\z}^2$. 
We observe that the plasma frequency $\omega_\text{pl}$ acts as a mass term for the electromagnetic waves.

\paragraph{Electromagnetic radiation sourcing gravitational waves. } 
Einstein's equations in the aforementioned gauge are $\Box h_{\mu\nu} =\kappa^2 T_{\mu\nu}$ with $\kappa^2=16\pi G/c^4$ and $T_{\mu\nu}$ being the energy-momentum tensor. For a wave traveling in the 3-direction, we must focus on the transverse piece which --to zeroth order in the metric perturbation-- is given in terms of the Maxwell stress tensor $\sigma_{ij} = B_\text{tot}^i B_\text{tot}^j -\delta^{ij} B_\text{tot}^2/2 $ as  $T_{ij} =-\sigma_{ij}$. Hence
\begin{equation}
\Box 
\begin{pmatrix}
h^{11} & h^{12}\\
h^{12} & h^{22}
\end{pmatrix}
=   \kappa^2
\begin{pmatrix}
-\dfrac{1}{2} (B_\text{tot}^1)^2+\dfrac{1}{2} (B_\text{tot}^2)^2+\dfrac{1}{2} (B_\text{tot}^3)^2& -B_\text{tot}^1 B_\text{tot}^2 \\
 -B_\text{tot}^1 B_\text{tot}^2 & \dfrac{1}{2} (B_\text{tot}^1)^2-\dfrac{1}{2} (B_\text{tot}^2)^2+\dfrac{1}{2} (B_\text{tot}^3)^2 \\
\end{pmatrix}\,.
\end{equation}
Here $B^i_\text{tot}$ includes the contributions from the external magnetic field and the electromagnetic wave. In the previous equation, we can neglect the quadratic piece in $A^{\mu}$  as well as the term  quadratic in the external magnetic field, which can not source gravitational waves. Under these assumptions and $\partial_i = (0,0,\partial_{\z})$ 
\begin{equation}
\Box 
\begin{pmatrix}
h^{11} & h^{12}\\
h^{12} & h^{22}
\end{pmatrix}
=   \kappa^2
\begin{pmatrix}
 -\partial_{\z} A^2 B^1-\partial_{\z} A^1 B^2 & \partial_{\z} A^1 B^1-\partial_{\z} A^2 B^2 \\
 \partial_{\z} A^1 B^1-\partial_{\z} A^2 B^2 & \partial_{\z} A^2 B^1+\partial_{\z} A^1 B^2 \\
\end{pmatrix}\,.
\label{eq:heq0}
\end{equation}
Eqs.~\eqref{eq:Aeq0} and \eqref{eq:heq0} explicitly show that the component of the external magnetic field parallel to the direction of motion of the waves does not affect their propagation. Furthermore, given this situation, without loss of generality, one can take $(B^1,B^2) = (B,0)$ by performing a rotation on the transverse plane. In that reference frame, we find 
\begin{align}
\left(\Box+\omega^2_\text{pl}/c^2 \right)A_\lambda = 
-B\,
\partial_{\z} h_\lambda
&&
\text{and}
&&
\Box 
 h_\lambda
=   \kappa^2 B\,
\partial_{\z} A_\lambda \,,
\label{eq:waveeqfull}
\end{align}
where $h_\times = h^{12}$ and $A_\times = A^1$ as well as $h_+ = -h^{22}$ and $A_+ =-A^2$.   Assuming a plane wave of frequency $\omega$, traveling in the positive direction with $\omega \geq \omegapl$, 
the exact solution of Eqs.~\eqref{eq:waveeqfull} can be written as 
\begin{align}
\hspace{-8pt}
\psi(t,\z)  \equiv
\begin{pmatrix}
\sqrt{\mu}\,\, A_\lambda \\
\frac{1}{\kappa}\, h_\lambda
\end{pmatrix}
=
e^{-i\omega (t-t_0)}
e^{i K (\z-\z_0) }
\psi(t_0,\z_0)\,,\,
&&
\text{with}
&&
K = 
\begin{pmatrix}
 \frac{ \mu}{c}\sqrt{\omega^2+\left(\frac{\kappa B}{1+\mu}\right)^2} \,  &  -i\frac{\sqrt\mu\,\kappa B }{1+\mu}\\
i \frac{\sqrt\mu\, \kappa B }{1+\mu} &  \frac{1}{c}\sqrt{\omega^2+\left(\frac{\kappa B}{1+\mu}\right)^2} 
\end{pmatrix}
 \,.
\label{eq:sol1}
\end{align}
Here  
$\mu= \sqrt{1-\omega^2/\omegapl^2}$ is the refractive index of the electromagnetic waves in the absence of the Gertsenshtein effect, which is assumed to be uniform. Notice that the eigenvalues of $K$, denoted by $k_\gamma$ and $k_g$, are the wave numbers of the resulting oscillation modes, whose corresponding group velocities to order ${\cal O} ((\kappa B)^2)$ satisfy
\begin{equation}
1-\frac{v_g}{c} =\frac{v_\gamma}{\mu c}-1=\frac{c^2\kappa^2 B^2}{2\omegapl^2}=  10^{-46}  \left(\frac{B}{\unit[1]{nG}}\,\frac{\unit[1]{ Hz}}{\omegapl}\right)^2 \,.
\end{equation}
Given the current constraint on this quantity of order $ 10^{-16}$ from neutron-star mergers~\cite{Monitor:2017mdv}, the effect of cosmological magnetic fields on the GW velocity is negligible.  
The magnitude of the off-diagonal element of $e^{-i\omega (t-t_0)} e^{i K (\z-\z_0)}$ in Eq.~\eqref{eq:sol1} determines the conversion probability of gravitational waves into electromagnetic radiation and vice-versa. To calculate this, we  define the oscillation length, $\losc = 2/(k_\gamma-k_g)$,   and note that the exponential matrix is given by
\begin{align}
e^{i K (\z-\z_0)} = e^{\frac{i}{2} \text{tr}{K}(\z-\z_0)} \left(\cos\left(\frac{\z-\z_0}{\losc}\right)  {1\!\!1} + i \losc \sin\left(\frac{\z-\z_0}{\losc}\right) \left( K -\frac{1}{2} \text{tr}{K}   {1\!\!1} \right)  \right).
\end{align}
The conversion probability is thus
 \begin{align}
P_\text{homogeneous}  =\left(|K_{12}| \,\losc\, \sin\left(\frac{\z-\z_0}{\losc}\right)\right)^2,
&&
\text{with}
&&
\losc^{-1}= \dfrac{1}{2}\sqrt{\left(\frac{\omega(1-\mu)}{c}\right)^2+ \kappa^2 B^2}\,.
\label{eq:Puniform}
\end{align}
Typically $\losc \ll \z-\z_0$ and the probability averages to
\begin{equation}
 P_\text{homogeneous}= \dfrac{1}{2}   \losc^2 |K_{12}|^2\,,
\label{eq:Puniformav}
\end{equation}
which is in particular independent of $\z - \z_0$.

\subsection{The effect of inhomogeneities and the magnetic-field power spectrum}

Inhomogeneities in the magnetic field and the electron density  make the coefficients in the wave equation position-dependent.
Since the oscillation lengths we consider in this work are significantly smaller than the scale of such position-dependent effects, we can still use Eq.~\eqref{eq:sol1} by adding the conversion probabilities in different patches where the magnetic field and the electron density are uniform. We now justify this procedure and prove that it leads to a boost factor in the conversion probability of Eq.~\eqref{eq:Puniformav}. 

In the presence of the inhomogeneities, the solution of the wave equation can be cast as  
\begin{align}
\psi(t,\z) =  e^{-i \omega (t-t_0)}  {\cal U}(\z,\z_0)\psi (t_0, \z_0) \,,
&&
\text{and}
&&
\partial_{\z}  {\cal U}(\z,\z_0) \approx i K(\z) {\cal U}(\z,\z_0)\,,
\label{eq:sol2}
\end{align}
with $K(\z)$ defined as in Eq.~\eqref{eq:sol1}. The fact that $[K(\z),K(\z')]\neq0$ prevents us from writing ${\cal U}(\z,\z_0)$ in exponential form, as we did above. Nonetheless, we can perturbatively solve for it  in terms of the magnetic field. More precisely, as argued in the main text, the oscillation length is dominated by the plasma term, which allows to neglect terms quadratic in $B$ in $K(\z)$. Then, we can split $K(\z)$ in a $B$-independent  piece and  a part linear in $B$,
\begin{align}
K(\z) = K_0(\z)+\delta K(\z) \,, &&\text{with} && K_0(\z) = 
\begin{pmatrix}
K_{11}(\z) &0\\
0 & K_{22}(\z)
\end{pmatrix}
 \,,
&&\text{and}
&&
\delta K(\z)= K_{12}(\z)
\begin{pmatrix}
0 &1\\
-1 & 0
\end{pmatrix}
 \,.
\end{align}
To this end, notice  that  $[K_0(\z),K_0(\z')]=0$ and that the second part of Eq.~\eqref{eq:sol2} implies that 
\begin{equation}
\partial_{\z} \left(e^{-i \int^{\z}_{\z_0} d\z' K_0(\z')} {\cal U}(\z,\z_0) \right) = i e^{-i \int^{\z}_{\z_0} d\z' K_0(\z')} \delta K (\z) {\cal U}(\z,\z_0)\,,
\end{equation}
which --after integration-- leads to
\begin{eqnarray}
 e^{-i \int^{\z}_{\z_0} d\z' K_0(\z')} {\cal U}(\z,\z_0)  &=& {1\!\!1}+ i \int^{\z}_{\z_0} d\z' e^{-i \int^{\z'}_{\z_0} d\z'' K_0(\z'')} \delta K(\z')  {\cal U}(\z',\z_0)\nonumber \\
&= & {1\!\!1}+ i \int^{\z}_{\z_0} d\z' e^{-i \int^{\z'}_{\z_0} d\z'' K_0(\z'')} \delta K(\z') e^{i \int^{\z'}_{\z_0} d\z'' K_0(\z'')} +  {\cal O}\left((\kappa B(\z-\z_0))^2\right)  \,.
\end{eqnarray}
In particular, noting that $\losc(\z)^{-1} = (K_{11}(\z)-K_{22}(\z))/2+{\cal O}(\delta K^2)$, we can cast the magnitude of the off-diagonal element as 
\begin{equation}
|{\cal U }_{12}(\z,\z_0) | = \bigg|\int^{\z}_{\z_0} d\z' e^{-2i \int^{\z'}_{\z_0} d\z'' \losc(\z'')^{-1} }  K_{12}(\z') \bigg| +  {\cal O}\left((\kappa B(\z-\z_0))^2\right) \,.
\end{equation}
The conversion probability is then
\begin{equation}
P = |{\cal U}_{12} (\z,\z_0)|^2 = \int^{\z}_{\z_0} d\z'\int^{\z}_{\z_0}d\tilde{\z}'  e^{-2i \int^{\z'}_{\tilde{\z}'} d\z'' \losc(\z'')^{-1} } K_{12}(\z')  K_{12}(\tilde{\z}')^*+  {\cal O}\left((\kappa B(\z-\z_0))^3\right) \,,
\label{eq:Pnonuniform0}
\end{equation}
which exactly reduces to Eq.~\eqref{eq:Puniform}, when $K$ is $\z-$independent. Cosmological homogeneity and isotropy requires
~\cite{Durrer:2013pga} 
\begin{equation}
\langle \mathbf{B}_i(\mathbf{x}) \mathbf{B}_j(\mathbf{x}')\rangle = \frac{1}{(2\pi)^3 a(t)^4} \int d^3 k e^{i \mathbf{k}\cdot (\mathbf{x}'-\mathbf{x})} \left(\left(\delta_{ij}-\hat{k}_i \hat{k}_j\right) P_B(k) -i  \epsilon_{ijk} \hat{k}_k  P_{aB}(k)\right)\,,
\label{eq:PB}
\end{equation}
Notice that the adiabatic evolution of the magnetic field due to cosmic expansion is determined by the scale factor $a(t)$.  We are interested in the transverse component of the magnetic field, $\sum^2_{i=1}\langle \mathbf{B}_i(\mathbf{x}) \mathbf{B}_i(\mathbf{x}')\rangle$, and therefore  the antisymmetric spectrum, $P_{aB}$, is not relevant here. On the other hand, 
the average magnetic field is $\langle B^2 \rangle =1/(\pi^2 a(t)^4) \int^\infty_0 dk k^2 P_B(k)  $, which can be used to define the magnetic field at a particular scale $\lambda$ as~\cite{Durrer:2013pga}
\begin{align}
 \langle B^2 \rangle  = \int_{-\infty}^\infty d\log\lambda \, B_\lambda^2  \,,&& \text{with}&& B_\lambda^2 \equiv \frac{8\pi}{\lambda^3 a(t)^4}P_B\left(\frac{2\pi}{\lambda}\right) \,,
\label{eq:coherencel}
\end{align} 
as well as the coherence length
\begin{equation}
\lambda_B = \int_0^\infty d \lambda \frac{B_\lambda^2}{\langle B^2 \rangle }   \,.
\end{equation}
Eq.~\eqref{eq:PB} also leads to
\begin{equation}
\langle B(\z)B(\z')\rangle = \frac{1}{(2\pi)^3 a(t)^4} \int d^3 k  e^{i \mathbf{k}\cdot \mathbf{\hat{z}} (\z'-\z)} P_B(k)  (1+ \mathbf{\hat{k}\cdot \mathbf{\hat{z}}}) 
\end{equation}
 During the dark ages, the electron density is expected to have inhomogeneities similar to those of dark matter, which are relevant at distances of the order of the horizon size during matter-radiation equality~\cite{Venhlovska:2008uc,Dvorkin:2013cga,2012PhRvD..85d3522F}. In contrast,  the magnetic field may have inhomogeneities on much smaller scales~\cite{Durrer:2013pga}, as suggested in Fig.~2 of the main text. Consequently and for simplicity, we first assume that the electron density is uniform. A posteriori, the treatment of the magnetic-field inhomogeneities will indicate how to deal with those of the electron density.

Furthermore, recall that the magnetic field can be safely neglected in $\losc$ and it only enters in Eq.~\eqref{eq:Pnonuniform0} through the expression $K_{12} (\z') K_{12} (\tilde{\z}')\propto B(\z') B(\tilde{\z}')$. We will also assume that $\z-\z_0$ is sufficiently small so that the Universe expansion can be ignored, then $\losc$ is essentially constant. 
Under these assumptions, the averaged probability from Eq.~\eqref{eq:Pnonuniform0}  is  
\begin{eqnarray}
\langle P\rangle &\simeq& \frac{1}{(2\pi)^3 a(t)^4} \frac{\langle |K_{12}|^2\rangle}{\langle B^2\rangle}\int d^3k  \int^{\z}_{\z_0} d\z'\int^{\z}_{\z_0}d\tilde{\z}'  e^{ i \left( 2\losc^{-1}+ \mathbf{k}\cdot \mathbf{\hat{z}} \right)\left(\tilde{\z'}-\z'\right)}  P_B(k)  (1+ \mathbf{\hat{k}\cdot \mathbf{\hat{z}}})
\nonumber\\
&=& \frac{(\z-\z_0)^2}{(2\pi)^3 a(t)^4} \frac{\langle |K_{12}|^2\rangle}{\langle B^2\rangle}\int d^3k \,\text{sinc}^2\left( \left(\losc^{-1}+ \frac{1}{2} \mathbf{k}\cdot \mathbf{\hat{z}}\right) (\z-\z_0) \right)  P_B(k)  (1+ \mathbf{\hat{k}\cdot \mathbf{\hat{z}}})
\nonumber\\
&\to& 
 \frac{(\z-\z_0)^2}{(2\pi)^3 a(t)^4} \frac{\langle |K_{12}|^2\rangle}{\langle B^2\rangle}
\int^\infty_0 2\pi k^2 dk   
\int^{1}_{-1} d(\cos \theta)\,
\pi
\delta\left( \left(\losc^{-1}+ \frac{1}{2}k \cos\theta\right) (\z-\z_0) \right)  P_B(k)  (1+k\cos \theta)\,,
\end{eqnarray}
where sinc$(x)\equiv \sin x/x$. In the last line we use the fact that $\losc\ll \z-\z_0$, for which the sinc vanishes unless its argument is zero. More precisely, for $q\gg 1$,  sinc$^2(q x)\to \pi \delta(q x) $. Due to this delta function,  positive values of $\cos\theta$ do not contribute to the integral while the negative ones ensure that $k>2\losc^{-1}$, leading to 
\begin{eqnarray}
\langle P\rangle &=&
 \frac{(\z-\z_0)^2}{(2\pi)^3 a(t)^4} \frac{\langle |K_{12}|^2\rangle}{\langle B^2\rangle}
\int^\infty_{2 \losc^{-1}} 2\pi k^2 dk   
\,
\left(
 \frac{2 \pi}{k (\z-\z_0) }
\,P_B(k)  \left(1+ \frac{2}{k\, \losc }\right)
\right)\nonumber\\
&=&
 \frac{2\pi\,(\z-\z_0)}{ a(t)^4} \frac{\langle |K_{12}|^2\rangle}{\langle B^2\rangle}
\int^{\pi \losc }_0 \frac{ d\lambda}{\lambda^3}   
\,
\,P_B\left(\frac{2\pi}{\lambda}\right)  \left(1+ \frac{\lambda}{\pi \losc }
\right)\nonumber\\
&=&
 \frac{1}{ 2 } (\z-\z_0) \langle |K_{12}|^2\rangle
\int^{\pi \losc }_0 d\lambda   
\,\frac{B^2_\lambda}{\langle B^2\rangle}  \left(\frac{1}{2}+ \frac{\lambda}{2\pi \, \losc }\right)
\,.
\end{eqnarray}
We thus find
\begin{align}
\langle P\rangle 
= \frac{(\z-\z_0)\, \cal F}{\losc} \langle P_\text{homogeneous} \rangle
\,,&&\text{with}&& {\cal F}=
\frac{1}{ \losc}\int^{\pi \losc }_0 d\lambda \,   
\,\frac{B^2_\lambda}{\langle B^2\rangle}  \left(\frac{1}{2}+ \frac{\lambda}{2\pi \, \losc }\right) =  \frac{\pi(1+\xi)}{2} \frac{B^2_{\xi\pi \losc}}{\langle B^2\rangle} 
\,,
\end{align}
where the integral has been evaluated using the mean-value theorem and therefore $0<\xi<1$. 
The formula on the left has been mentioned in Ref.~\cite{Cillis:1996qy} without specifying the model-dependent factor ${\cal F}$.  The rate is thus
\begin{equation}
\langle\Gamma_{g\leftrightarrow\gamma}\rangle = \frac{c\langle P\rangle}{\z-\z_0 } =\frac{ \cal F}{c\,\losc} \langle P_\text{homogeneous}\rangle  \,.
\label{eq:rateAppendix}
\end{equation}
 From Eq.~\eqref{eq:coherencel}, it is clear that ${\cal F}\gtrsim 1$ is impossible. 
Furthermore, on small scales, $\lambda < \pi \ell_\text{osc}$, the power spectrum is expected to decrease as a power law $P_B(k) \propto k^{-\alpha}$, or equivalently, $B_\lambda^2/\langle B^2 \rangle \sim \left(\lambda/\lambda_B\right)^{\alpha-3}$~\cite{Durrer:2013pga}. We thus expect ${\cal F}\sim (\losc/\lambda_B)^{\alpha-3}$.  A scale-invariant power spectrum gives $\alpha = 3 \rightarrow B_\lambda^2 =\text{cte} \rightarrow {\cal F} \sim 1$. This is unlikely because at small scales there is a damping  of the power induced by magnetohydrodynamical effects. In fact, in the light of this, the more
 realistic (and conservative) scenario corresponds to $\alpha\sim 4$~\cite{Durrer:2013pga}, or ${\cal F}\sim \losc/\lambda_B$. According to Eq.~\eqref{eq:rateAppendix}, this gives a rate  given by Eq.~(5) of the main text with the inhomogeneity scale $\Delta \z=\lambda_B$. 

Accounting for inhomogeneities in the electron density is analogous. Following a similar procedure (see also~\cite{Carlson:1994yqa}), we also obtain Eq.~\eqref{eq:rateAppendix} with ${\cal F}$ now depending on the power spectrum associated with $n_e$. We expect the latter to track the dark matter as argued above. In the main text we conservatively take Eq.~(5)
with  $\Delta \z= \text{min}[\lambda_\text{EQ}, \lambda_B]$, where $\lambda_{EQ} $ is the scale of the matte perturbations set by  matter-radiation equality.

\subsection{The Boltzmann-equation approach}
The Boltzmann equation describing the GW and CMB distribution is 
\begin{align}
\hat{L} f_{\gamma/g} =\pm \langle\Gamma_{g\leftrightarrow\gamma}\rangle (f_g -f_\gamma)\,,
\end{align}
with $\hat{L} \equiv\partial_t - H \omega \partial_\omega = - H \left( T \partial_T + \omega \partial_\omega \right)$.
Note that the sum of both distributions simply redshifts 
 because $\hat{L}(f_\gamma+f_g) = 0$, which implies that  $ f_\gamma(\omega,T)+f_g(\omega,T)= f_\gamma(T_\ini\omega/T,T_\ini)+f_g(T_\ini\omega/T,T_\ini)$. 
On the other hand, $\Delta  \equiv (f_\gamma-f_g)/2$ satisfies 
\begin{eqnarray}
\hat{L}  \Delta &=& - 2\langle\Gamma_{g\leftrightarrow\gamma}\rangle \Delta\,,
\end{eqnarray} 
which can be solved for fixed values of $\omega/T$ as  $ \Delta (\omega,T) =\Delta(T_\ini\omega/T ,T)\exp \left( -2 \int^{T_\ini}_{T} \frac{dT'}{T'H(T')}\langle\Gamma_{g\leftrightarrow\gamma}\rangle|_{\omega/T' =\text{cte}}\right)$. Thus, the full solution is 
\begin{align}
\begin{bmatrix} f_\gamma(\omega,T)\\ f_g(\omega,T) \end{bmatrix} =e^{-\intf} \begin{bmatrix} \cosh \intf & \sinh \intf \\  \sinh \intf& \cosh \intf \end{bmatrix}\begin{bmatrix} f_\gamma(\frac{T_\ini}{T}\omega,T_\ini) \\ f_g(\frac{T_\ini}{T}\omega,T_\ini)\end{bmatrix}\,,
&&
\text{with} 
&&
\intf = \int^{T_\ini}_{T} \frac{\langle\Gamma_{g\leftrightarrow\gamma}\rangle|_{\omega/T' =\text{cte}} }{T'\, H(T')} dT'\,,
\end{align} 
which reduces to Eq.~(9) of the main text for $\intf \ll 1$. Note that this $\intf$ is the same as the line-of-sight integral (hence the condition $\omega/T' =\text{cte}$) in Eq.~(6).

\subsection{The density-matrix approach}

Eq.~\eqref{eq:sol2} determines the evolution of pure states describing electromagnetic and gravitational radiation coupled by means of the Gertsenshtein effect. Nonetheless, accounting for decoherence effects requires to go beyond pure states by considering statistical mixtures. Such mixed states are elegantly described by a density matrix.  Using the states in Eq.~\eqref{eq:sol2}, we can define
\begin{equation}
\rho \left(t_0, \z_0\right) = {\cal N} \left( f_\gamma^0 |\gamma\rangle\langle \gamma| +f_g^0|g\rangle \langle g | \right)\,
\label{eq:rho0}
\end{equation}
where ${\cal N}$ is a normalization constant chosen so that Tr~$\rho =1$. As long as decoherence effects are absent the evolution is unitary and, according to Eq.~\eqref{eq:sol2}, we have
\begin{equation}
\rho(t,\z) = {\cal U}(\z,\z_0) \rho(t_0, \z_0) {\cal U}(\z,\z_0)^\dagger\,.
\label{eq:ev}
\end{equation}
The diagonal entries determine $f_\gamma$ and $f_g$, while non-zero off-diagonal elements --also called coherences-- indicate interference between $|\gamma\rangle$ and $|g\rangle$ and thus coherent evolution. In particular,
\begin{eqnarray}
f_\gamma (t,\z)& =& |{\cal U}_{11}(\z,\z_0)|^2 f^0_\gamma +  |{\cal U}_{12}(\z,\z_0)|^2 f^0_g =  f^0_\gamma+  |{\cal U}_{12}(\z,\z_0)|^2 \left(f^0_g-f^0_\gamma\right)\,,\\
f_g (t,\z)& =& |{\cal U}_{22}(\z,\z_0)|^2 f^0_g +  |{\cal U}_{21}(\z,\z_0)|^2 f^0_\gamma =  f^0_g+  |{\cal U}_{21}(\z,\z_0)|^2 \left(f^0_\gamma-f^0_g\right)\,. 
\end{eqnarray} 
Inhomogeneities in the electron density or the magnetic field induce decoherence making the off-diagonal entries vanish, which leads to a density matrix such as that in Eq.~\eqref{eq:rho0}. In that case one can still use Eq.~\eqref{eq:ev} to determine $\rho$ in patches where its evolution is coherent, i.e. $\rho(t_{i+1},\z_{i+1}) = {\cal U}(\z_{i+1},\z_{i}) \rho(t_{i},\z_{i} ) {\cal U}(\z_{i+1},\z_{i})^\dagger,$ as long as $|\z_{i+1}-\z_i|$ is much smaller than the scale of the inhomogeneities. Performing a telescopic sum under the assumption that  $|{\cal U}_{12} (\z_{i+1},\z_{i})|\ll 1$, we find
\begin{eqnarray}
f_\gamma(t,\z)  &=&f^0_\gamma+ (f^0_g-f^0_\gamma) \sum_i  |{\cal U}_{12}(\z_{i+1},\z_i)|^2 \nonumber\\&\to& f^0_\gamma + (f^0_g-f^0_\gamma)\int d\z   \left( \lim_{\z_{i+1}\to \z_i} \frac{|{\cal U}_{12}(\z_{i+1},\z_i)|^2}{\z_{i+1}-\z_i}\right)=f^0_\gamma + (f^0_g-f^0_\gamma) \int \langle \Gamma_{g\leftrightarrow\gamma}(\z_i)  \rangle \,dt \,,
\end{eqnarray}
which again reduces to Eq.~(9) of the main text, when trajectories along the line-of-sight are considered.

\end{widetext}